\documentclass[prl,preprint,aps]{revtex4}
\begin{document}
\title{Comment on: Breakdown of Bohr's Correspondence Principle \\
by: Bo Gao Phys. Rev. Lett. 83, 4225 (1999).}
\author{C. Tannous}
\affiliation{Laboratoire de Magn\'{e}tisme de Bretagne, UPRES A CNRS 6135,
Universit\'{e} de Bretagne Occidentale, BP: 809 Brest CEDEX, 29285 FRANCE}
\author{J. Langlois}
\affiliation{ Laboratoire des Collisions Electroniques et Atomiques,
Universit\'{e} de Bretagne Occidentale, BP: 809 Brest CEDEX, 29285 FRANCE}
\date{September 1, 2000}


\maketitle

Gao applied LeRoy and Bernstein ~\cite{Leroy70} semi-classical analysis for the energy
levels in a potential of the form $-C_n/r^n$ to sequences of scaled energy differences 
(SED) progressing towards low lying states and found a better agreement with the 
semi-classical prediction for low-lying levels until eventually 
interactions having shorter range come into play. The sequences given
in his Table 1 are stopped before the agreement deteriorates. He claimed that 
Bohr's Correspondence Principle breaks down for all quantum systems in which 
the asymptotic interaction is of the form $1/r^n$ with $n>2$.\\

We checked that for the energy levels obtained by Stwalley et al. ~\cite{Stwalley78}
with the same potential ~\cite{Movre77}, the agreement with the semi-classical
approximation is better for higher vibrational quantum numbers in 
agreement with Bohr's correspondence principle. \\

\begin{table}[htbp]
\begin{center}
\begin{tabular}{|c|c|c|c|c|}
\hline
Index    &   Stwalley et al. &  SED  & CFM & SED \\
\hline
1&-1.7887&&-1.7864488& \\
2&-1.5617&0.9566&-1.5595638&0.9571\\
3&-1.3566&0.9698&-1.3546072&0.9702\\
4&-1.1723&0.9820&-1.1703990&0.9827\\
5&-1.0075&0.9939&-1.0057071&0.9946\\
6&-0.86087&1.0057&-0.8592631&1.0059\\
7&-0.73125&1.0159&-0.7297908&1.0164\\
8&-0.61729&1.0259&-0.6159534&1.0265\\
9&-0.51770&1.0352&-0.5164882&1.0358\\
10&-0.43120&1.0440&-0.4301217&1.0444\\
11&-0.35657&1.0520&-0.3556148&1.0525\\
12&-0.29261&1.0595&-0.2917693&1.0601\\
13&-0.23820&1.0662&-0.2374560&1.0669\\
14&-0.19224&1.0727&-0.1916002&1.0732\\
15&-0.15374&1.0783&-0.1531893&1.0789\\
16&-0.12176&1.0832&-0.1212854&1.0840\\
17&-9.5438(-02)&1.0873&-9.5022588(-02)&1.0886\\
18&-7.3940(-02)&1.0929&-7.3608452(-02)&1.0926\\
19&-5.6599(-02)&1.0957&-5.6325744(-02)&1.0962\\
20&-4.2754(-02)&1.0988&-4.2530867(-02)&1.0994\\
21&-3.1831(-02)&1.1014&-3.1650591(-02)&1.1021\\
22&-2.3323(-02)&1.1039&-2.3180420(-02)&1.1044\\
23&-1.6791(-02)&1.1058&-1.6679756(-02)&1.1064\\
24&-1.1854(-02)&1.1075&-1.1768655(-02)&1.1081\\
25&-8.1873(-03)&1.1088&-8.1228816(-03)&1.1094\\
26&-5.5165(-03)&1.1100&-5.4689541(-03)&1.1106\\
27&-3.6136(-03)&1.1109&-3.5793742(-03)&1.1115\\
28&-2.2916(-03)&1.1116&-2.2676168(-03)&1.1122\\
29&-1.3995(-03)&1.1122&-1.3831806(-03)&1.1128\\
30&-8.1747(-04)&1.1128&-8.0683859(-04)&1.1133\\
31&-4.5276(-04)&1.1132&-4.4613249(-04)&1.1138\\
32&-2.3503(-04)&1.1135&-2.3110168(-04)&1.1142\\
33&-1.1252(-04)&1.1141&-1.1035443(-04)&1.1146\\
34&-4.8564(-05)&1.1146&-4.7468345(-05)&1.1152\\
35&-1.8262(-05)&1.1153&-1.7767388(-05)&1.1160\\
36&-5.6648(-06)&1.1165&-5.4747950(-06)&1.1172\\
37&-1.3175(-06)&1.1185&-1.2597092(-06)&1.1194\\
38&-1.9247(-07)&1.1148&-1.2716754(-07)&1.2814\\
39& -1.1215(-08)& 1.1131&  & \\
40& -4.1916(-11)& 1.1131&  & \\

\hline     
\end{tabular}
\end{center}
\end{table}

Clearly a disagreement must exist between the energy levels calculated by
Gao and those by Stwalley et al. for the same potential.\\ 

Thus we set out to evaluate directly the full spectrum of the Movre-Pichler ~\cite{Movre77}
potential with a powerful tunable accuracy method that embodies a control of accuracy of the
quantum eigenvalues. It is based on the Canonical Function Method (CFM) ~\cite{Tannous99} that
 allows us to evaluate eigenvalues close to the ground state as well as close to highly excited states near
the continuum. Hence the semi-classical approximation
can be tested with high accuracy close to the continuum limit as well as 
at lower energies.\\

We used for the $0^-_g$ electronic state of the $^{23}{Na}_2$ molecule the same parameters
as those of Stwalley et al. ~\cite{Stwalley78} who found 37 energy levels and extrapolated 3 extra ones. 
Our results along with Stwalley et al.'s (in $cm^{-1}$) and the corresponding SED are
displayed in the Table progressing from the Ground up. The semi-classical approximation is better for higher
 vibrational quantum numbers as seen in the table (as compared to Gao's table 1) in our case and  Stwalley 
et al.'s, except for our very last level (number 38) that is quite close to dissociation.\\

The fact that two entirely different and independent methods reached the same result leads us to
believe that no breakdown of the Bohr's correspondence principle has been clearly established yet, in the above work.\\

{\bf Acknowledgements}: Part of this work was performed on the IDRIS machines at the CNRS (Orsay).\\

\end{document}